\documentclass[]{interact}

\usepackage{epstopdf}
\usepackage[caption=false]{subfig}

\usepackage[natbibapa,nodoi]{apacite}
\renewcommand\bibliographytypesize{\fontsize{10}{12}\selectfont}

\theoremstyle{plain}

\theoremstyle{definition}

\theoremstyle{remark}

\usepackage{longtable}
\usepackage{amsmath, amssymb, amsfonts}
\usepackage{algorithmic}
\usepackage[ruled,vlined]{algorithm2e}
\usepackage{xcolor}
\usepackage{graphicx}
\usepackage{multirow}
\usepackage{rotating}

\begin{document}


\title{GCNIDS: Graph Convolutional Network-Based Intrusion Detection System for CAN Bus}

\author{
\name{Maloy Kumar Devnath \thanks{Corresponding Author: Maloy Kumar Devnath Email: maloyd1@umbc.edu} }
\affil{Department of Information Systems, University of Maryland, Baltimore County (USA)}
}

\maketitle

\begin{abstract}
The Controller Area Network (CAN) bus is a standard protocol used for communication between various electronic control units (ECUs) in modern vehicles. However, it has been demonstrated that the CAN bus is vulnerable to remote attacks, which can compromise the safety and functionality of the vehicle. To address this issue, intrusion detection systems (IDSs) have been proposed to detect and prevent such attacks. In this paper, we propose a novel approach for intrusion detection in the CAN bus using a Graph Convolutional Network (GCN)~\citep{zhang2019graph}. By leveraging the power of deep learning, we can achieve higher accuracy in detecting attacks while minimizing the need for feature engineering. Our experimental results demonstrate that the proposed GCN-based approach outperforms state-of-the-art IDSs in terms of accuracy, precision, and recall. We also show that the GCN-based approach can effectively detect mixed attacks, which are more challenging to detect than single attacks. Moreover, the proposed approach requires less feature engineering and is more suitable for real-time detection systems. To the best of our knowledge, this is the first work that applies the GCN to CAN data for intrusion detection. Our proposed approach can significantly enhance the security and safety of modern vehicles by detecting attacks and preventing them from compromising the functionality of the vehicle.
\end{abstract}


\section{Introduction}

\label{sec:intro}

A standard for communication between the many modules that make up the electrical system of
a vehicle is known as the Controller Area Network bus or CAN bus for short. The CAN bus is
used in Tesla vehicles to connect many systems and components~\cite{nie2017free,zniti2023hardware}. Some of these systems and
components include the powertrain, the battery, the sensors (lidar, mmWave radar), and the displays~\cite{spencer2021design,devnath2023systematic}. Through the
sending and receiving of digital messages, the CAN bus makes it possible for various
components to communicate with one another. Because of this, the various systems can
collaborate and share information with one another to guarantee that the vehicle functions in the
most effective manner. For instance, the powertrain control module can receive information
about the level of charge from the battery management system over the CAN bus. Based on this
information, the module can then change the amount of power that is produced. In general, the
CAN bus is a crucial component of the electrical architecture of the Tesla vehicle since it enables
communication and coordination between all of the different components in an effortless manner.
Researchers have already demonstrated remote attacks on crucial electronic control units (ECUs)
for vehicles by leveraging controller area networks (CANs)~\cite{zniti2023hardware,wei2023real}. The CAN bus is also a potential target for attackers, who may exploit vulnerabilities in the system to launch attacks on electronic control units (ECUs). Such attacks can compromise the security and safety of the vehicle, leading to potential harm to passengers and property. Therefore, maintaining the security of the CAN bus is of utmost importance in ensuring the safety and reliability of modern vehicles. Several security measures have been proposed to mitigate the risks associated with CAN bus attacks, including intrusion detection systems (IDSs), encryption techniques, and access control mechanisms. IDSs are particularly important in detecting and preventing attacks, as they can identify malicious activities in the network and alert the appropriate authorities to take action~\cite{farag2017cantrack}. However, IDSs are often limited in their ability to detect mixed attacks, in which multiple types of attacks are launched simultaneously or sequentially. Therefore, there is a need for more robust and generalizable IDSs that can detect a wide range of attacks in real time. In addition, present intrusion detection systems (IDSs) frequently promise to defend against a certain kind of attack, which may leave a system open to countless other types of attacks. A generalizable intrusion detection
system (IDS) that is able to recognize a wide variety of assaults in the shortest amount of time
possible has more practical utility than an attack-specific IDS, which is not an easy undertaking
to complete successfully. In this work~\cite{refat2022detecting}, the researchers have used the graph properties as a feature for applying machine learning techniques. They also have detected mix attacks which are a combination of DoS, Fuzzy, and Spoofing but the most difficult one is the replay attack, they do not consider the replay attacks in mixed attacks. They have to conduct a lot of feature engineering in order to make the machine learning model work, which can make it more difficult to make decisions during real-time
identification. It would be fantastic for autonomous systems if we could take a step that would
require less feature engineering but still have a high level of accuracy. We are going to look into
the possibility of a mixed attack which is a combination of DoS, Fuzzy, Spoof, and Replay attack  because we are aware that, at the present moment, we do not
know which attack will take place. In the event that a DOS assault takes place, it will not reveal
that I am a DOS attacker. Any kind of attack could take place at any moment. Therefore, if we
think about attacking the CAN bus in a mixed (any time any attacks can happen) fashion, we will have more success. Our proposed model shows that it works better than the state art work~\cite{refat2022detecting} for the mixed (combination of DoS, Fuzzy, and Spoof attacks).

In this project, our aim is to tackle the following research problems with a comprehensive and scholarly approach:
\begin{itemize}
    \item We want to minimize the amount of feature engineering than the existing work~\cite{refat2022detecting} so that it can be readily included in a real-time detection system. 
    \item In order to make the primary protocol more useful in the real-time detection system, one of our goals is to improve the accuracy of other and mixed attacks (a combination of DoS, Fuzzy, Spoof, and Replay attacks) without having to change the protocol itself which is not considered in the existing study~\cite{refat2022detecting}.
    \item To the best of our knowledge, it is first to apply the Graph Convolutional Network using only two graph-based features maximum indegree and maximum outdegree on graph-based CAN data~\cite{rahman2017basic,zhang2019graph} and Our proposed GCN model has yielded superior results compared to the existing methodology~\cite{refat2022detecting}.
\end{itemize}

The study is organized as follows - Section~\ref{sec:literature} discusses recent works on CAN bus security challenges and Graph-based anomaly detection.  Data description and processing techniques are described in Section~\ref{sec:dataset}. The model training mechanism is elaborated in Section~\ref{sec:method}. Section~\ref{sec:experiments} reports the experimental results and findings, and Section~\ref{sec:conclusion} concludes the study.

\section{Related Works}
\label{sec:literature}

Koscher et al. are first able to demonstrate that an attacker who can get access to virtually any ECU can go through a broad array of safety-critical systems by directly interfacing with the OBD-IIport~\cite{experimental}.They have full control of a wide range of functions: disabling the brakes, stopping the engine, and controlling other vehicle functions by using reverse engineering code. Checkoway et al. later demonstrate that a vehicle can be accessed remotely~\cite{comprehensive}. Earlier research has shown that vehicles are insecure within internal networks. They have gained successful access non-physically. They have attacked the redtooth and infotainment systems of vehicles. Miller and Valasek have analyzed the rate of messages for in-vehicle network intrusion detection~\cite{ASR}.It should be possible to detect anomalous messages by analyzing the distribution rate of messages. Valasek and Miller have demonstrated that it is possible to have real-world attacks on multiple vehicles by using the CAN bus~\cite{miller2015remote}.  The brakes of a Jeep Cherokee are successful while it is on a live highway. Moore et al. have proposed that the regularity of CAN message frequency detects the anomaly~\cite{MIS}. A similar detection method has been proposed by Gmiden says that Moore’s detector relies on the time intervals of CAN messages~\cite {AID}. Regularity in the signal frequencies is seen by them, and from there they hypothesize that accurate detection of regular-frequency signal injection attacks is possible by monitoring the inter-signal wait times of CAN bus traffic will provide. Zhou et al. present an advanced CAN bus anomaly detection system for intelligent vehicles by integrating DNN technology and a triple loss network~\cite{ADCB}. Firstly, the methodology first extracts data features as a set of vectors by using the deep network, and then calculates the similarity between two real-time extracted CAN data sequences. From there, the triple loss uses another calibrated data sequence to find out the abnormal data. They use only malicious data. Verendel et al. have proposed a honeypot security mechanism that has been placed at the wireless gateway acting as a decoy in simulating the in-vehicle network~\cite{AAT}. Attacking information is collected and analyzed to update the later version of the system. The most challenging in deploying a honeypot is that it must be realistic as possible. The attacker should not have information about that. Wolf et al. have proposed an architecture that is based on firewall signatures for securing vehicular communication gateways~\cite{ESI}. It filters authorized controllers to exchange valid messages. However, they have also said that it could not fully shield the vehicle network as most modern vehicles have interfaces that enable access to the entire car system. Marchetti et al. propose the first algorithm based on the analysis of the sequences of messages which flow on the CAN bus~\cite{ADO}. Without knowing the message specifications, the feature can be extracted from the CAN messages. The computational requirements of the proposed algorithm are low enough with compatible with fewer hardware resources. Kang et al. propose an intrusion detection system that is based on the deep neural network (DNN) to secure the CAN network~\cite{IDS}. After reducing high-dimensional CAN packet data, the method figures out the underlying statistical properties of normal and attack packets. Then, identify the attack after extracting the corresponding features. Graph-based anomaly detection is not a new idea. Paudel et al. use the publicly available tool graph-based anomaly detection tool (GBAD) (Eberle and Holder 2007)~\cite{DTO}. It is a graph-based approach proposed by them. They demonstrate that GBAD not only focuses on anomalies within an entity but also it allows us to find the anomalies that exist in an entity’s relationship. The authors introduce a novel approach for graph-based anomaly detection by adding background knowledge to the evaluation metrics used in a traditional graph-mining approach~\cite{NGB}. Background knowledge is added in the form of rule coverage reporting the percentage of the final graph covered by the instances of the substructure. The authors hypothesize that by assigning negative weights to the rule coverage, they can discover anomalous substructures.  Velampalli et al. uses a graph-based approach that analyzes the data for suspicious employee activities at Kasios~\cite{DSP}. Graph-based approaches are so much power to handle rich contextual data and provide a deeper understanding of data due to the ability to discover patterns in databases that are not easily found using traditional query or statistical tools. They focus on graph-based knowledge discovery in structural data to mine for interesting patterns and anomalies. Paudel et al. have proposed a sketching approach in a graph stream called SNAPSKETCH~\cite{snapsketch}. From a biased-random walk, a simplified hashing of the discriminative shingles is generated that is used for anomaly detection in the dynamic graphs. Eberle et al. propose a novel graph-based anomaly detection approach named Graph-based Outlier Detection by representing home IoT traffic as a real-time graph stream~\cite{DDA}. They detect DoS attacks in real-time by processing graph data efficiently. Tanksale et al. propose an intrusion detection system based on Support Vector Machine that can detect anomalous behavior with high accuracy~\cite{tanksale:2019}. They also give a process for selecting parameters and features. The drawback of their works is that they only consider dos attacks. Song et al. propose an Intrusion Detection System to secure the CAN bus from cyber-attacks which is based on a deep convolutional neural network (DCNN)~\cite{song:2020}. They make a frame builder that converts the CAN bus data into a grid-like structure so that CAN bus data can be fitted to the DCNN. The drawback of their works is that they do not consider replay and mixed attacks. Adding computers in automobiles has brought several benefits such as driver comfort, vehicle efficiency, and performance. Along with these kinds of benefits, the dependence of automobiles on such devices has increased potential attacks and threats to mankind.
So many authors agree that the CAN network lacks security. Some critics think that it has not been built with security in mind.  It does not provide any security against malicious attacks~\cite{ueda2015security2},~\cite{studnia2013survey3},
~\cite{carsten2015system7},~\cite{boudguiga2016simple8},~\cite{staggs2013hack13} and~\cite{hoppe2011security14}.
Hiroshi et al. tells that in-vehicle networks message spoofing is considered one of the main threats as it is possible to take control of a critical safety systems display
a falsified value to a vehicle’s maintainer~\cite{ueda2015security2}.
Being not capable of distinguishing between a legitimate ECU and a malicious one, replay attacks are made in the CAN network. As an unauthorized device can be easily connected to the CAN-Bus, it is possible to transmit spoof messages~\cite{ueda2015security2},~\cite{foster2015exploring4}, ~\cite{carsten2015system7},~\cite{staggs2013hack13} and~\cite{hoppe2011security14}.
Moreover, the CAN network is also not freed from Denial of Service and fuzzy attacks. DoS attacks can be performed by sending high-priority messages again and again by assigning successive dominant bits on the bus~\cite{staggs2013hack13},~\cite{foster2015exploring4},~\cite{boudguiga2016simple8} and~\cite{hoppe2011security14}. It is possible for an attacker to cause fuzzy attacks that can inject messages of randomly spoofed identifiers having arbitrary data. For this reason, receiving lots of functional messages can cause unintended vehicle behaviors which may cause damage to human life~\cite{candata}.

Machine learning or deep learning algorithm has been applied in multiple applications~\cite{sarker2021machine,protikuzzaman2020predicting,devnath2023systematic,anwar2023heteroedge}. Following these different machine learning models in recent years, various approaches have been proposed to detect attacks on Controller Area Network (CAN) bus along with anomaly detection~\cite{DDA,snapsketch,DSP}, a crucial component in modern vehicles. Some of these methods relied on feature engineering~\cite{refat2022detecting,maloy2022ggnb}, while others utilized deep learning models to classify CAN bus attacks~\cite{song:2020}. Some of them consider only DoS attacks~\cite{tanksale:2019}. However, there remains room for improvement in terms of accuracy and reducing the reliance on feature engineering. In this study, we present a graph-based approach that incorporates only two features, namely, the maximum indegree and maximum outdegree of nodes in the graph. Our approach leverages Graph Convolutional Networks (GCN) to achieve reliable accuracy in detecting mixed attacks on the CAN bus, a problem that has not been adequately addressed by previous research. We demonstrate the effectiveness of our method using real-life CAN bus data and compare our results with those obtained from other state-of-the-art techniques. Our findings show that our approach outperforms existing methods in terms of accuracy and reduces the need for extensive feature engineering.

\section{Dataset}
\label{sec:dataset}

\subsection{Dataset Description}
The OTIDS (Operational Technology Intrusion Detection System) dataset is a publicly available dataset of CAN bus traffic designed specifically for intrusion detection research in operational technology (OT) environments. The dataset has been collected by researchers from the University of Twente and the Netherlands Organization for Applied Scientific Research (TNO). The dataset contains CAN bus traffic captured from a real-world OT environment, specifically a moving vehicle. The data has been collected using a CAN bus logger in real-time while the vehicle is in operation. The data has been preprocessed to remove any duplicate or corrupt packets. The dataset contains 5 sessions of CAN bus traffic, each with a duration of approximately 1 hour. Each session contains thousands of CAN bus packets, with a total of over 8 million packets across all sessions. The packets are labeled as either normal or anomalous based on their behavior, making the dataset suitable for both normal and anomaly detection research. The OTIDS dataset also includes metadata, such as the timestamp, ID, and data field of each packet. This metadata can be used to extract features and train machine-learning models for intrusion detection. Overall, the OTIDS dataset is a valuable resource for researchers interested in developing intrusion detection systems for OT environments, particularly those using CAN bus networks.

The OTIDS dataset contains various types of attacks on the CAN bus network, with different message sizes. Here are some examples:

\begin{itemize}
    \item Denial of Service (DoS) attack: In this attack, the attacker floods the network with a large number of messages, causing the network to become congested and unresponsive. The messages in this attack are typically small in size, usually less than 8 bytes~\cite{candata}.
    \item Spoofing attack: In this attack, the attacker sends messages with forged source addresses, making it appear as if the messages are coming from a legitimate source. The messages in this attack are typically small in size, similar to normal messages on the network~\cite{candata}.

    \item  Fuzzing attack: In this attack, the attacker sends messages with intentionally malformed data to try to crash or exploit a vulnerable device on the network. The messages in this attack can vary in size, depending on the specific payload used~\cite{candata}.


    \item Replay attack: An replay attack is a type of cyber attack in which an attacker pretends to be someone else in order to gain access to sensitive information or resources. This type of attack is often used in phishing and social engineering attacks, where the attacker impersonates a trusted individual or entity to trick the victim into divulging sensitive information or performing an action that is harmful to the victim or their organization~\cite{candata}.
    
\end{itemize}

Overall, the message sizes in the OTIDS dataset vary widely depending on the type of attack being performed. It's important to note that the message sizes alone may not be sufficient for detecting attacks, and other features such as message frequency and content may also need to be considered. For our case, we have considered DoS, Fuzzy, Spoof, Replay, and a mix of all four attacks.

The number of messages is considered in our experiment shown in Table~\ref{tab:messages}.

\begin{table}[!h]
\center
\caption{Number of attacked and attack-free messages.}

\label{tab:messages}

\begin{tabular}{cc}
\hline
Attack type & Number of messages\\ 
\hline
DoS attack & 656,579 \\ 
Fuzzy attack & 591,990 \\
Spoofing attack & 500,900 \\
Replay attack & 995,472 \\ 
Attack free state & 2,369,868 \\ \hline
\end{tabular}
\end{table}
\subsection{CAN Data Frame:}

CAN bus is developed by Robert Bosch in 1986~\cite{van2011canauth}. It is a broadcast type of bus which is a message-based protocol. Here, no host is required. It maintains serial half-duplex asynchronous communication using two differential communication. Earlier, it was designed for automobiles. Later, it was used in other perspectives. In the CAN data frame, we have five fields. They are arbitration, control, Data CRC, and  ACK field. Along with these, there must be a Start of frame bit and an End of frame bit. Now we go through the one-by-one component of the data frame which is shown in Table~\ref{tab:can_data}.

\begin{table}
\caption{CAN bus dataframe}
\label{tab:can_data}
\center
\begin{tabular}{@{}ccc@{}}
\hline
Field & Size & Description \\ \hline
Start of frame & 1 bit & Indicates the beginning of a new frame \\
Arbitration & 12 bits & Identifies the message priority and sender ID \\
Control & 6 bits & Contains the message control information \\
Data & 0-8 bytes & Carries the actual message data \\
CRC & 16 bits & Provides error detection for the message \\
Ack & 2 bits & Indicates message received successfully or not\\ 
End of frame & 7 bits & Indicates the end of the frame \\ \hline
\end{tabular}
\end{table}

\begin{itemize}
    \item SOF- Start of frame  (1 bit). It indicates the start of a new frame in CAN network~\cite{hpl2002introduction}.
    \item Arbitration field – Here, an 11 bits message identifier and a Remote Transmission Request bit. It is used to set the priority of the data frame at the time arbitration process~\cite{hpl2002introduction}. RTR defines whether the frame is a data frame or a remote frame. It is for 1 bit.  
    \item Control field – It is user-defined. IDE bit means identifier extension. A dominant IDE bit indicates 11 bits standard frame identifier~\cite{hpl2002introduction}. Recessive IDE bit indicates an extended 29 bits identifier. Then, in the control field, we have a data length code (4 bits) that defines the length of the data in the data field.
    \item Data field- We can have a maximum of 8 bytes in the data field. 4 bits of DLC actually control how many bytes of data will be available in the data field~\cite{hpl2002introduction}. It is user-defined.
    \item CRC field - Cyclic Redundancy Check field consists of 15 bits for Error detection during transmission. CRC will be computed by the sender before sending the frame. After receiving, the receiver will again compute CRC~\cite{hpl2002introduction}. The receiver will generate the error frame if the CRC does not match.
    \item ACK filed- There are two bits in the Acknowledge field as ACK and ACK delimiter bit~\cite{hpl2002introduction}. After receiving a valid message normally, a node replaces the ACK part with a dominant bit which was a recessive bit.
    \item EOF – CAN frame is ended by seven recessive bits~\cite{hpl2002introduction}.
\end{itemize}

\begin{table}
\center
\caption{Raw CAN bus data}
\label{tab:raw_can_data}
\center
\begin{tabular}{cccc}
\hline
Timestamp & Arbitration ID & DLC & Data   \\
\hline
1478198376 & 0316 & 8 & 05 21 68 09 21 21 00 6f   \\
1478198376 & 018f & 8 & fe 5b 00 00 00 3c 00 00   \\
1478198376 & 0260 & 8 & 19 21 22 30 08 8e 6d 3a  \\
1478198376 & 02a0 & 8 & 64 00 9a 1d 97 02 bd 00  \\
1478198376 & 0329 & 8 & 40 bb 7f 14 11 20 00 14  \\
\hline
\end{tabular}
\end{table}

\subsection{Dataset Processing}

The proposed IDS aims to enhance the security of the CAN bus communication system by using deep learning analysis as a basis for detecting anomalies. We are going through several steps:

\begin{itemize}
    \item The first step is to transform the CAN bus messages into a more meaningful graph structure using graph theory~\cite{refat2022detecting,maloy2022ggnb}. This is achieved by dividing the CAN bus messages which are shown in Table~\ref{tab:raw_can_data} into a number of windows and deriving the relationships among all the arbitration IDs for each window. Graphs are a popular method for indicating relationships among data that are too complicated to express using simple text or other forms of data structure. As graph theory can represent complex relationships of data in a very simple manner, the proposed IDS leverages this to represent CAN bus data windows in a meaningful structure.

    \item The algorithm constructs graphs for each window of 200 messages and returns the overall constructed graph lists~\cite{refat2022detecting,maloy2022ggnb}. The algorithm initializes all necessary variables and computes the total number of messages in the given CAN dataset. It then uses a loop to iterate over every CAN bus message from the dataset, extracting the adjacent CAN messages and their corresponding IDs. The algorithm constructs an adjacency list from the arbitration IDs extracted from two sequential CAN messages.
    
\end{itemize}

\section{Methodology}
\label{sec:method}

\begin{figure*}
  \centering
  \hfill
  \subfloat []{\includegraphics[width=0.48\textwidth]{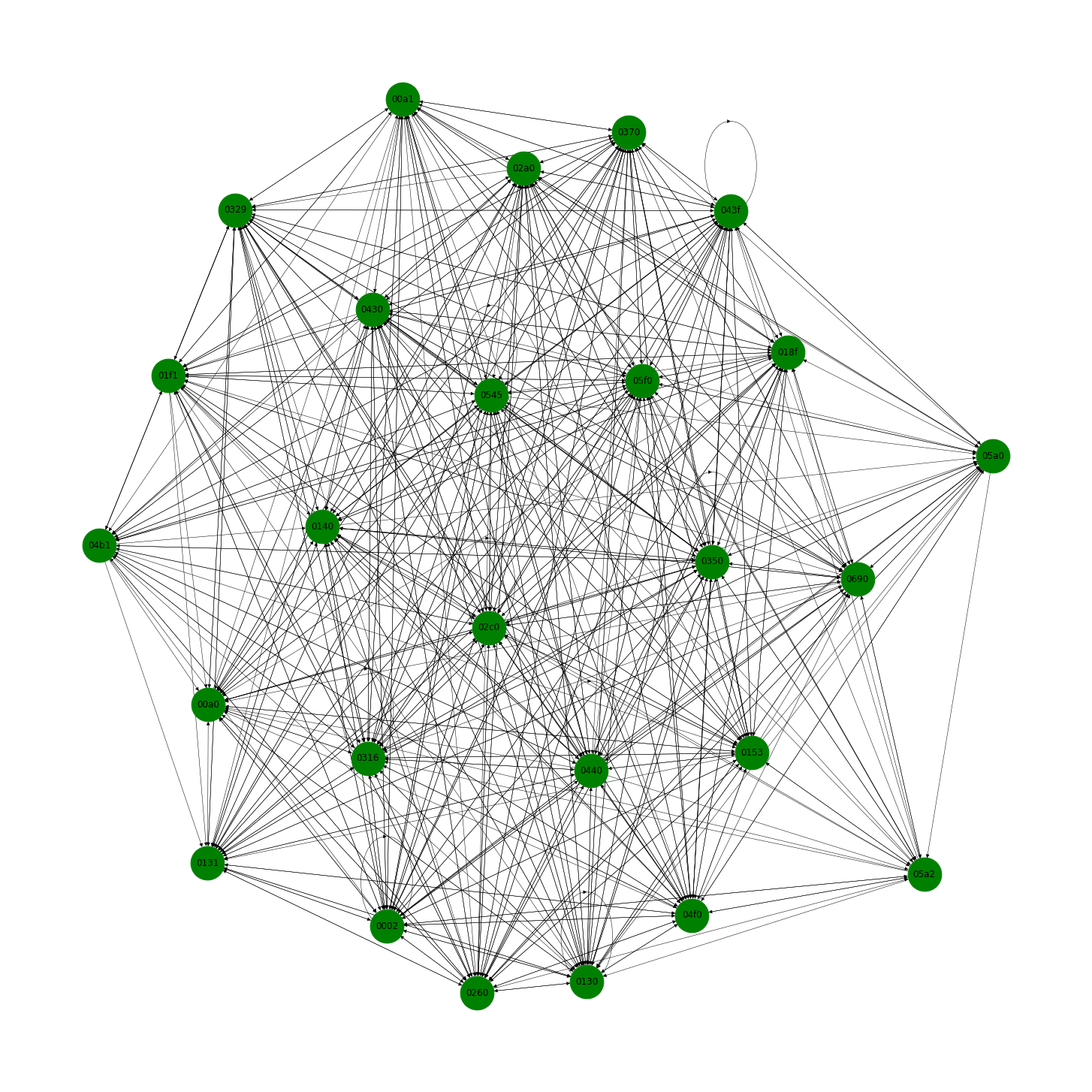}}
  \hfill
  \subfloat []{\includegraphics[width=0.48\textwidth]{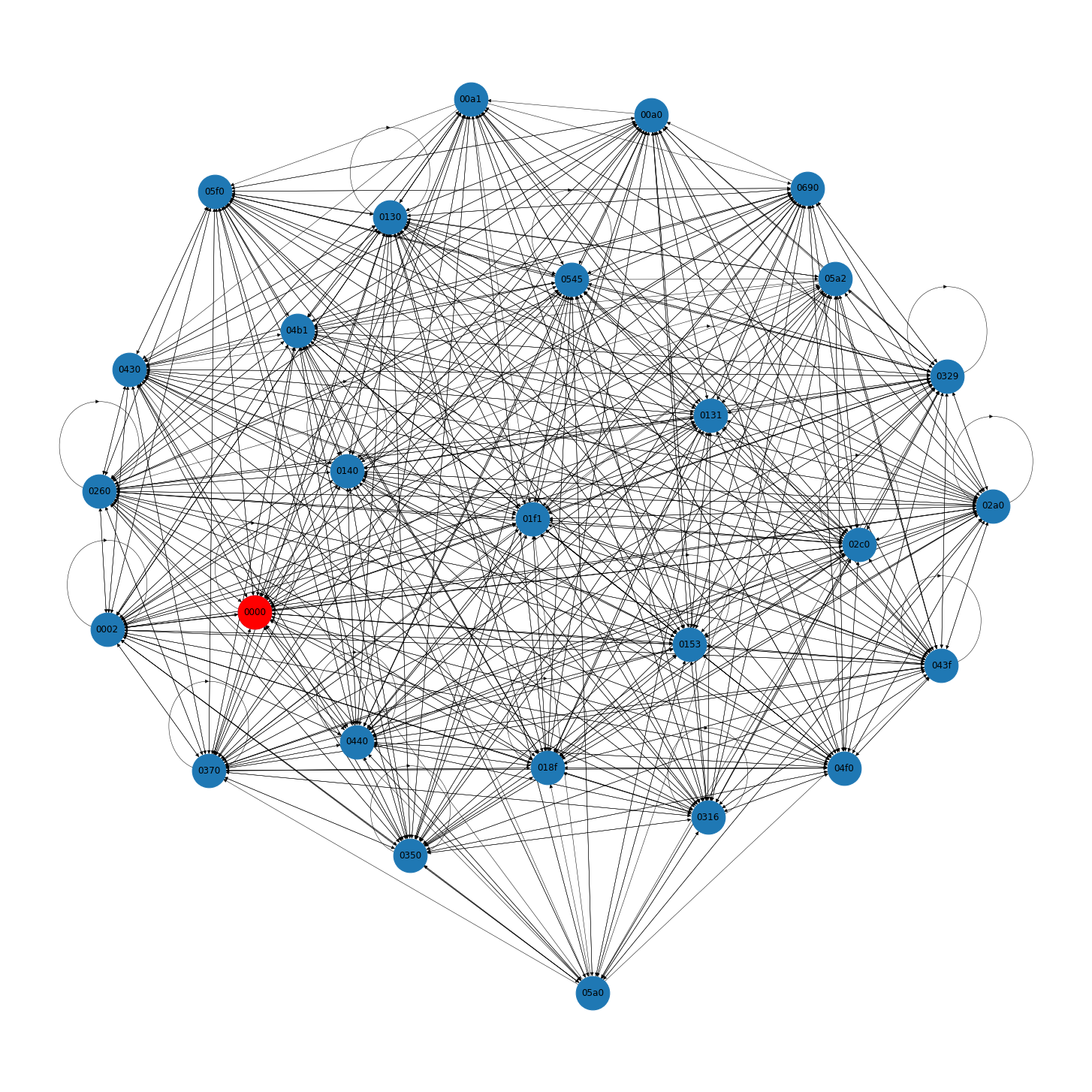}}
  \hfill
  \caption{(a) Attack-free graph where all the attack-free nodes (message ids) are shown in green color.  (b)Attacked graph (DoS-attacked graphs) where attacked node (message id) is shown in red color.}
  \label{fig:attackfree_attacked graph}

\end{figure*}

The proposed intrusion detection methodology in this study utilizes a graph-based approach that incorporates statistical analysis for detecting anomalies in Controller Area Network (CAN) bus communication system. The algorithm consists of several steps, starting with the construction of a graph from CAN bus messages using the following equation:
\begin{equation}
G = (V, E)
\end{equation}
where $G$ represents the constructed graph, $V$ represents the nodes, and $E$ represents the edges. Each node $v_i$ in the graph represents the arbitration ID of a CAN bus message, and each edge $e_{i,j}$ represents a sequential relationship between two adjacent CAN messages. The attacked and attack-free graphs are shown in Figure~\ref{fig:attackfree_attacked graph}. In Figure~\ref{fig:attackfree_attacked graph}(b), the red node indicates the attacked message id.

The graph-based features are extracted from the constructed graph. We consider the indegree and outdegree of the nodes as the features of the node. The GCN model takes the node features as input and propagates them through the graph structure to obtain the final embeddings~\cite{zhang2019graph}. The node features are represented as a feature matrix $X$ with dimensions $n \times f$, where $n$ is the number of nodes in the graph and $f$ is the number of features per node.

The basic graph convolution operation can be defined as:

\begin{equation}
    H = f(A,X) = \sigma(AXW)
\end{equation}

where $A$ is the adjacency matrix of the graph, $X$ is the node feature matrix, $W$ is the weight matrix, $\sigma$ is the activation function, and $H$ is the new node feature matrix after convolution.

Now, we can define the GCN as a series of graph convolution layers:
\begin{equation}
    H^{(0)} = X
\end{equation}
\begin{equation}
    H^{(l+1)} = f(A,H^{(l)}) = \sigma(A H^{(l)} W^{(l)})
\end{equation}
for $l = 0, 1, \ldots, L-1$, where $L$ is the number of layers.

Here, $H^{(0)}$ is the initial node feature matrix, which is typically set to be the input feature matrix $X$. $H^{(l+1)}$ is the output of the $l$-th layer of the GCN, and it is computed by applying the graph convolution operation to the input feature matrix $H^{(l)}$. The weight matrix $W^{(l)}$ in each layer is learned during the training process, and it determines the linear transformation applied to the node features in that layer. The activation function $\sigma$ is applied element-wise to the output of the graph convolution operation, and it introduces non-linearity to the model.

Overall, the GCN equation can be seen as a way to learn node representations that capture both the local and global information of the graph structure, by applying a series of graph convolution operations to the input node features.

The method defines a GCN model with two graph convolution layers and a readout layer followed by a final classifier. The model takes as input the graph structure of CAN messages and extracts features from it to classify the input as anomalous or normal. The loss function used in the training of the GCN model is   binary cross-entropy loss shown in the following Equation.
\begin{equation}
    L = -\frac{1}{N} \sum_{i=1}^{N} [y_i \log(\hat{y_i}) + (1-y_i) \log(1-\hat{y_i})]
\end{equation}
where:
\begin{itemize}
    \item $L$ is the binary cross-entropy loss.
    \item $N$ is the total number of samples.
    \item $y_i$ is the true label of the $i$-th sample (either 0 or 1).
    \item $\hat{y_i}$ is the predicted probability of the $i$-th sample being in class 1 (i.e., the output of the model for the $i$-th sample)
    \item $\log$ is the natural logarithm function.
\end{itemize}
where $y_i$ is the true label of the $i$-th sample, $\hat{y}_i$ is the predicted probability of the $i$-th sample belonging to the positive class, and $N$ is the total number of samples. The loss function is minimized during the training process to optimize the model's parameters.
In the forward pass of the model, the input data (represented as a graph) is passed through the two graph convolutional layers. Each graph convolution layer takes as input the node features and the graph structure, represented as edge indices, and applies a linear transformation to generate new node embeddings. The first layer takes the initial node features (in this case, the node is the arbitration ID) and maps them to a hidden dimension of size $8$. The second layer takes the output of the first layer as input and generates a new set of node embeddings with the same hidden dimension.

After the graph convolution layers, the model applies a readout layer to obtain a single feature vector representing the entire graph. This is done by computing the mean of the node embeddings across all nodes in the graph. Finally, a linear classifier with two output nodes is used to classify the input graph as either normal or anomalous.The dropout function is applied after the readout layer, with a probability of 0.5, to reduce overfitting during training.The model's hyperparameter is shown in Table~\ref{tab:model}.

\begin{table}
\center
\caption{Hyperparameter of the proposed GCN model}
\label{tab:model}
\begin{tabular}{cc}
\hline
Hyperparameter & Value \\ \hline
GCNConv1 input channels & 2 \\
GCNConv1 output channels & 8 \\ 
GCNConv2 input channels & 8 \\ 
GCNConv2 output channels & 8 \\ 
Linear input size & 8 \\ 
Linear output size & 2 \\
Dropout probability & 0.5 \\ \hline
\end{tabular}
\end{table}

\section{Systematic Study}
\label{sec:experiments}

\begin{figure}
  \centering
  \subfloat []{\includegraphics[width=0.30\textwidth]{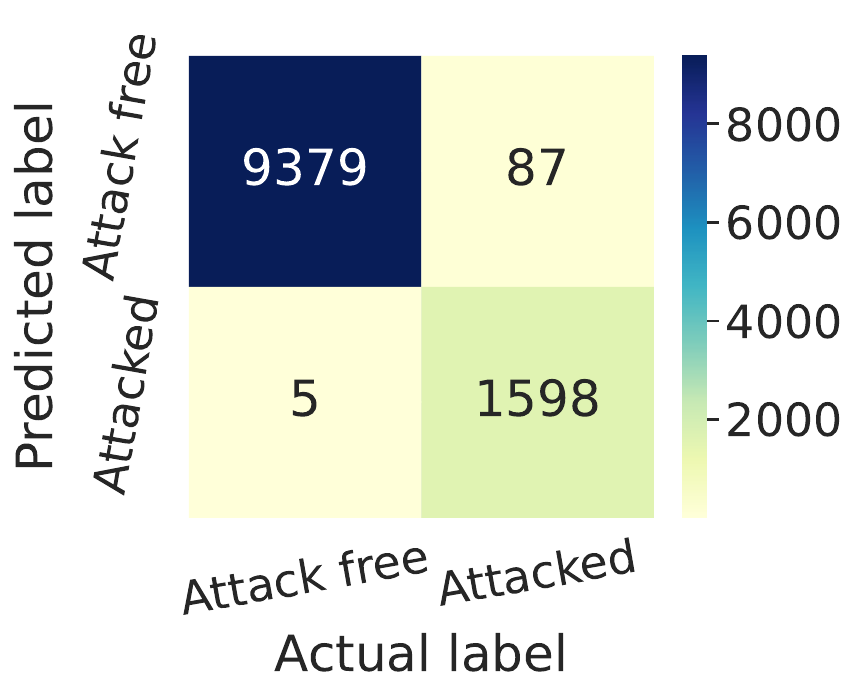}}
  \hfill
  \subfloat []{\includegraphics[width=0.30\textwidth]{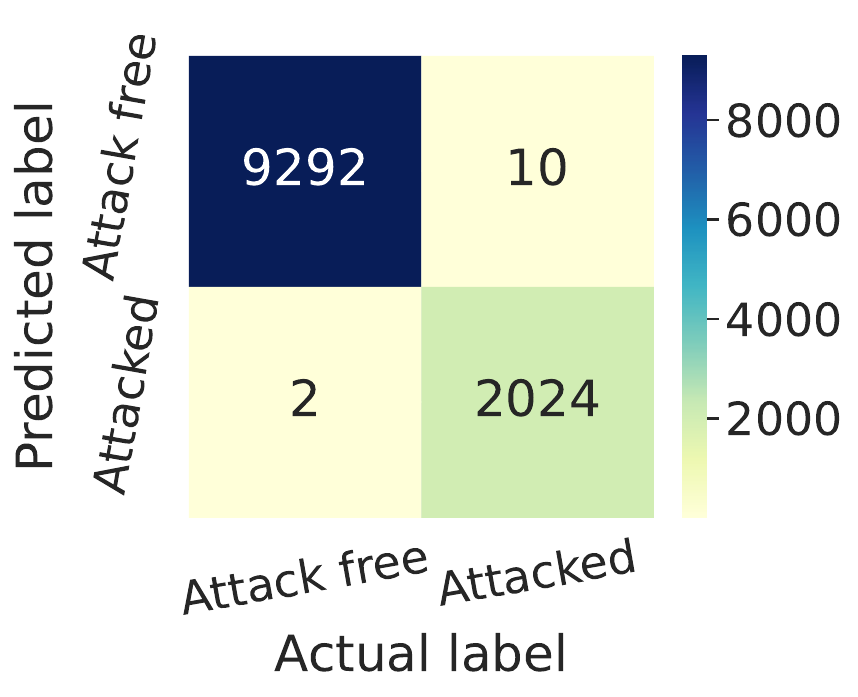}}
  \hfill
  \subfloat []{\includegraphics[width=0.30\textwidth]{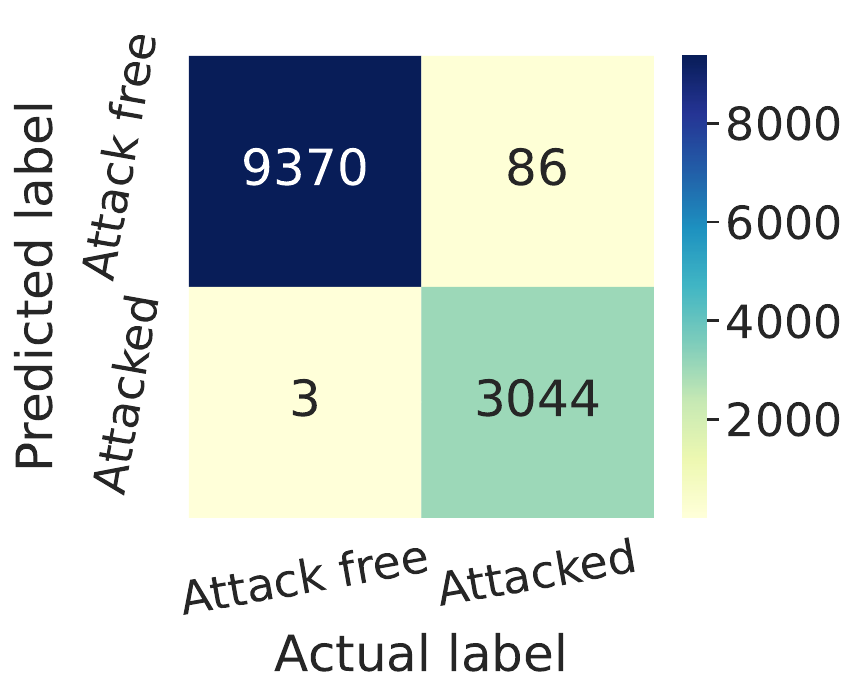}}
  \caption{(a)Dos attacked graphs detection: accuracy=0.9917 and wrong classification=0.0083 (b)Fuzzy attacked graphs detection: accuracy=0.9989 and   wrong classification=0.0011 (c)Spoof attacked graphs detection: accuracy=0.9929 and wrong classification=0.0071 }
  \label{fig:c_dfs}
\end{figure}
\begin{figure}
  \centering
  \subfloat []{\includegraphics[width=0.30\textwidth]{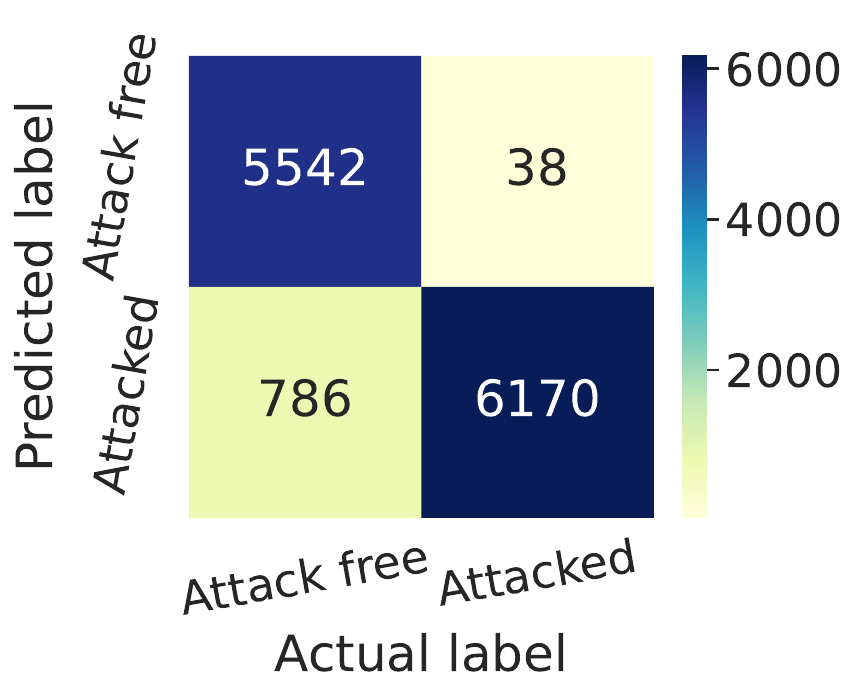}}
  \hfill
  \subfloat []{\includegraphics[width=0.30\textwidth]{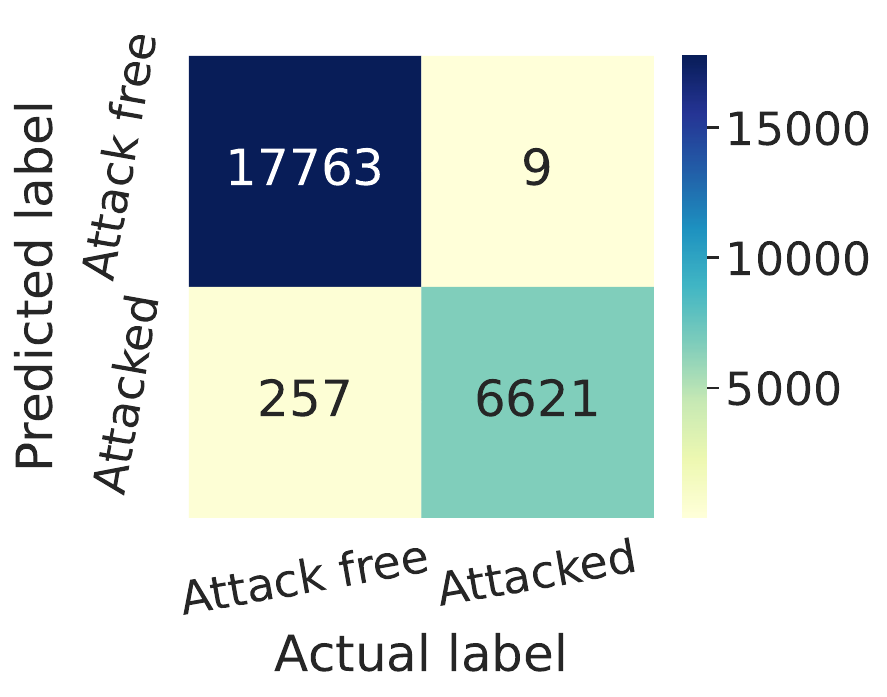}}
  \hfill
  \subfloat []{\includegraphics[width=0.30\textwidth]{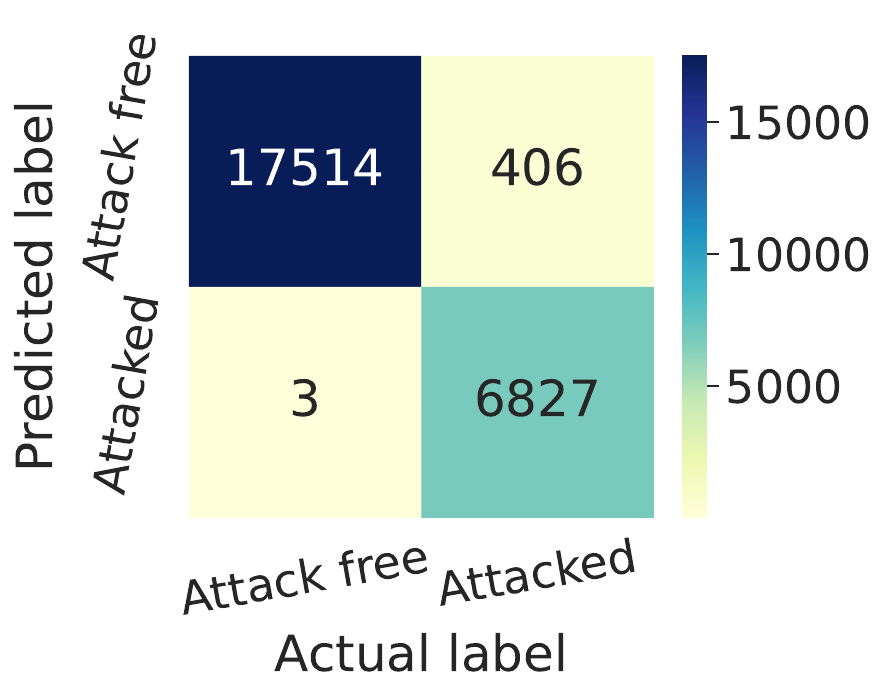}}
  \caption{(a) Replay attacked graphs detection:
accuracy=0.9343 and  wrong classification=0.0657 (b) Mixed attacked graphs (DOS, Fuzzy, Spoofing) detection: accuracy=0.9892 and  wrong classification=0.0108 (c) Mixed attacked graphs (DOS, Fuzzy, Spoofing, Replay) detection: accuracy=0.9835 and  wrong classification=0.0165 }
  \label{fig:c_rmo}

\end{figure}

For experimenting with our proposed methodology, we have used a real CAN dataset and performed analysis on NVIDIA GeForce RTX
3060 graphics card by using Python language. Firstly, we have collected the Raw CAN data which is described in this paper~\cite{candata}. Then, we have converted the Raw data into graph data as graphs can capture information from any relationship. Conversion of CAN data into graph data is based on a time window as CAN Data has a fixed message injection rate~\cite{maloy2022ggnb,refat2022detecting}. We have converted the CAN data into five different graphs data. We make the graph in such a way that these graphs are mutually independent. They are attack-free graphs, DoS-attacked graphs, fuzzy-attacked graphs, spoofing-attacked graphs, and replay-attacked graphs. Every graph is a collection of so many graphs. Like, $18565$ graphs are making attack-free graphs. Here, attack free graph is completely attacked-free. Otherwise, the rest of them are a combination of attack-free and attacked graphs. 

For detecting the DoS attacked graphs, we merge the attack-free and attacked graphs as attacked messages will not come distinctly from the attack-free messages in a real CAN messages scenario. For, detecting fuzzy, spoofing, and replay-attacked graphs, we also merge the attack free and attacked graphs. For designing a real-life CAN bus, we have to assume that any kind of attack may occur at any time. So, we merge all the attacked and attack-free graphs for detecting any kind of attack. It is named mixed attacks in the rest of the paper.

We have applied GCN to our graph data set which is a class of deep learning methods~\cite{zhang2019graph}. Generally, deep learning methods take more training time than other methods. We try to speed up the computation. As our graphs data set is usually small and we want to guarantee full GPU utilization, it may seem to us that it is a good idea to batch the graphs before inputting them into a Graph Neural Network. In the image domain, this method is typically done by rescaling or padding each example into a set of equally-sized shapes grouping the examples in an additional dimension~\cite{chua1998cnn}. But it is not feasible for our purposes. The above procedures may result in a lot of unnecessary memory consumption. PyTorch Geometric opts for another approach to achieve parallelization across numerous examples where adjacency matrices are stacked in a diagonal fashion~\cite{fey2019fast,foster2015exploring4}. It  makes a big graph that holds multiple isolated subgraphs and concatenates simple nodes and target features in the node dimension.

\begin{table}   
\tiny
\caption{When considering DoS and mixed attacks, the proposed methodology has better precision, recall, and F1 scores, respectively, than the state-of-the-art SVM classifier~\cite{tanksale:2019} and GBF~\cite{refat2022detecting}.}
\label{tab:big_table}
\center
\begin{tabular}{@{}lllllllllllll@{}}
\hline
Attack Type &  \multicolumn{3}{c}{~\cite{tanksale:2019}} &  \multicolumn{3}{c}{~\cite{song:2020}} &   \multicolumn{3}{c}{~\cite{refat2022detecting}} &  \multicolumn{3}{c}{GCNIDS (Proposed)}    \\ \hline

Evaluation metrics & Pr & Re & F1 & Pr & Re & F1 & Pr & Re & F1 & Pr & Re & F1 \\ \hline
DOS &0.46&0.89&0.61&1&1&1&1&1&1&0.99&1&1 \\ \hline
Fuzzy & - & - & - & 1 & 1 & 1 & 1 & 0.99 & 0.99 & 1 & 1 &1 \\ \hline
Spoofing & - & - & - & 1 & 1 & 1 & 0.98 & 0.94 & 0.96& 0.99 & 1 &1 \\ \hline
Replay & - & - & - & - & - & - & - & - & - & 0.99 & 0.88 & 0.93\\ \hline
Mixed (D, F, S) & - & - & - & - & - & - & 0.99 & 0.96 & 0.97 & 1 & 0.99 &0.99 \\ \hline
Mixed (D, F, S, R) & - & - & - & - & - & - & - & - & - & 0.98 & 1 &0.99 \\ \hline

\multicolumn{13}{l}{D means DoS, F means Fuzzy, S means Spoofing, R means Replay}
\end{tabular}

\end{table}



 

  

Along with the less training time apparently, we have some crucial advantages over other batching procedures as GCN operators rely on a message-passing scheme. As messages are not exchanged between two nodes that belong to distinct graphs, they do not need to be modified. There is no computational or memory overhead as adjacency matrices are saved sparsely. It  holds only non-zero entries like the edges. We will analyze our confusion matrix from Figure~\ref{fig:c_dfs}. First, we concentrate on the DoS attacked graphs which are shown in Figure~\ref{fig:c_dfs}(a). Here, we see that our proposed methodology wrongly detects 87 attack-free graphs which are attacked graphs. Our accuracy for detecting DoS-attacked graphs is quite satisfactory. It is 99\%. For detecting fuzzy attacked graphs, we observe that it does quite well. Figure~\ref{fig:c_dfs}(b) illustrates the scenario. It has been classified wrongly as only 0.11\%. The wrong classification numbers are so small only 12 overall from the test cases. Detecting spoof attacked graphs, we can that from Figure~\ref{fig:c_dfs}(c) it has classified incorrectly 86 attacked graphs as attack-free graphs from the overall  3130 attacked graphs. We know that replay-attacked graphs detection is one of the toughest attacks for detection. But our Figure~\ref{fig:c_rmo}(a) shows that our proposed methodology works quite well for detecting replay-attacked graphs. It only makes 38 wrong detections which attack free graphs in the test cases.   
We have already been informed that in real life mixed attacks will occur. We hope that it is our specialty to detect the mixed attack. For detecting the mixed attacks (only a combination of DoS, Fuzzy, and Spoof attacks), the confusion matrix~\ref{fig:c_rmo}(b) shows that we have good accuracy of 99\% which is better than state of art~\cite{refat2022detecting}. The proposed methodology can handle the detection of mixed attacks quite well. Apart from the detection of mixed attacks of the state of the art~\cite{refat2022detecting}, we also analyze the mixed attacks ( a combination of DoS, fuzzy, spoof, and replay attacks ) illustrated in Figure~\ref{fig:c_rmo}(c). We see that our proposed methodology is strong enough to detect any kind of attack.

From Table~\ref{tab:big_table}, we can see that our proposed methodology has better performance than SVM~\cite{tanksale:2019} classifier. In DCN~\cite{song:2020}, they are considering a couple of attacks but they do not consider the mixed attacks. But in our proposed methodology, we have considered mixed attacks which may occur in real life. Our proposed methodology even works better than the state of art graph-based features (GBF)~\cite{refat2022detecting} when we consider mixed attacks (only a combination of DoS, Fuzzy, and Spoof attacks).


\section{Discussion}
\label{sec:discuss}

We address the issues of developing and implementing a deep learning model that utilizes raw CAN data. These include challenges in real-time implementation with the label and without label data and the system's potential impact on daily life.

\begin{itemize}
    \item When we apply the GCN model, we consider the labeled dataset for training. If any unknown attack happens, the model will not capture it. It is the limitation of our model. We need to move to build a model in an unsupervised manner so that it can easily detect the abnormal behavior of the messages without the label.
    
    \item Integrating the intrusion detection pipeline into a real-time system can be straightforward, but there are various implementation challenges to consider, such as power consumption and unwanted messages from the surrounding auto-motives. These challenges present concrete obstacles to achieving real-time implementation.

    \item One limitation of our study is that we have used a single real-world data set for our experiments. It may be valuable to validate our methodology on other data sets to see how it generalizes to different CAN bus scenarios.

    \item In addition, while GCN is a powerful machine-learning technique, it can be computationally expensive, particularly for large data sets. In this study, we have tried to speed up the computation, but it is still an area for improvement. Future work may explore ways to further optimize the GCN-based approach to improve its scalability and reduce computation time.
     
\end{itemize}

\section{Conclusion}

\label{sec:conclusion}

In conclusion, we have proposed a novel GCN model which is a class of deep learning methods for detecting various types of attacks on CAN networks by converting the raw CAN data into graph data. Our experiments on a real CAN dataset show that our proposed methodology outperforms the state-of-the-art SVM classifier~\cite{tanksale:2019} and GBF~\cite{refat2022detecting} in terms of precision, recall, and F1 scores. We have also analyzed the performance of our methodology on other types of attacks, including fuzzy, spoofing, replay, and mixed attacks. Our methodology provides an effective and efficient solution for detecting different types of attacks on CAN networks, which is crucial for ensuring the security and reliability of automotive systems. Further research may be conducted to improve the real-time application of our methodology, as well as explore its application in other domains beyond automotive systems.

\section*{ACKNOWLEDGEMENT}
We want to thank Dr. Busiime Ngambeki for her collaboration on this project.

\bibliographystyle{apacite}
\bibliography{main.bib}

\end{document}